\documentclass[twocolumn,english,superscriptaddress]{revtex4-1}
\usepackage[T1]{fontenc}
\usepackage[latin9]{inputenc}
\usepackage{xcolor}
\usepackage{pdfcolmk}
\usepackage{amsmath}
\usepackage{amssymb}
\usepackage{graphicx}
\PassOptionsToPackage{normalem}{ulem}
\usepackage{ulem}

\makeatletter

\providecolor{lyxadded}{rgb}{0,0,1}
\providecolor{lyxdeleted}{rgb}{1,0,0}
\DeclareRobustCommand{\lyxadded}[3]{{\color{lyxadded}{}#3}}

\DeclareRobustCommand{\lyxsout}[1]{\ifx\\#1\else\sout{#1}\fi}

\makeatother

\usepackage{babel}
\begin{document}

\title{Geometry and kinetics determine the microstructure in arrested coalescence
of Pickering emulsion droplets}

\author{Zhaoyu Xie}

\affiliation{Department of Physics and Astronomy, Tufts University, 574 Boston
Avenue, Medford, MA 02155}

\author{Christopher J. Burke}

\affiliation{Department of Physics and Astronomy, Tufts University, 574 Boston
Avenue, Medford, MA 02155}

\author{Badel Mbanga}

\affiliation{Department of Physics and Astronomy, Tufts University, 574 Boston
Avenue, Medford, MA 02155}

\author{Patrick T. Spicer}

\affiliation{Complex Fluids Group, School of Chemical Engineering, UNSW Sydney,
Sydney, Australia}

\author{Timothy J. Atherton}
\email{timothy.atherton@tufts.edu}

\affiliation{Department of Physics and Astronomy, Tufts University, 574 Boston
Avenue, Medford, MA 02155}
\begin{abstract}
An important strategy to stabilize emulsions is to arrest coalescence
of the constituent droplets with an opposing rheological force. Colloidal
particles adsorbed on the surface of emulsion droplets in a Pickering
emulsion become increasingly crowded during successive coalescence
events because the combined surface area of coalescing droplets is
less than that of the constituent droplets. Beyond a critical density,
the particles form a rigid shell around the droplet and inhibit both
relaxation of the droplet shape and further coalescence. The resulting
droplets have a nonuniform distribution of curvature and, depending
on the initial coverage, may incorporate a region with negative Gaussian
curvature around the neck that bridges the two droplets. Here, we
resolve the relative influence of the curvature and the kinetic process
of arrest on the microstructure of the final state. Identifying the
dimensionless ratio of the rate of area change $\dot{A}$ to the diffusion
constant $D$ as a measure of the importance of kinetics in this system,
we show that this depends on the extrinsic geometry of the surface
as opposed to the static packings that depend solely on intrinsic
geometry. 
\end{abstract}
\maketitle

\section{Introduction}

Emulsions are mixtures of immiscible fluids where one fluid is dispersed
as droplets inside the other, continuous fluid. Such mixtures phase
separate or coarsen by \emph{coalescence}, fusion of droplets, and
\emph{Ostwald ripening}, exchange of fluid between droplets by diffusion
through the continuous phase. To stabilize an emulsion, it is therefore
necessary to inhibit these mechanisms. One strategy to do so is to
disperse micron sized or nanoscale particles in the mixture, forming
a Pickering emulsion\citep{pawar2011arrested,studart2009arrested,Dockx2018fx}.
The particles adsorb on the surface of the droplets, reducing the
interfacial contact between the two fluids. When two such droplets
come into contact, and the particle coverage fraction is less than
some critical value $\Phi_{c}$, the droplets fully coalesce. Because
the combined droplet has a lower surface area than the precursor droplets,
however, the coverage fraction necessarily increases. Conversely,
if droplets come into contact with a particle coverage that exceeds
the critical coverage fraction, the particles become crowded and prevent
the droplets from fully relaxing. The end result is typically a non-spherical
droplet with a solid interfacial shell that prevents further coalescence
of additional droplets and also tends to inhibit Ostwald ripening.
This phenomenon is an example of \emph{arrested coalescence}, which
can also be achieved by other offsetting rheological resistance such
as internal viscoelastic fluids\citep{pawar2012arrested}. 

Prior studies\citep{pawar2011arrested} of arrested coalescence have
shown that the point of arrest can be predicted by a simple geometric
model incorporating $\Phi_{0}$, the particle coverage fraction of
the precursor droplets prior to coalescence, and the relative size
of the two droplets. Once the expected coverage fraction for the combined
droplet would exceed $\Phi_{c}=\pi/\sqrt{12}\approx0.9$, the value
of random close packing in 2D, the coalescence will be arrested; if
$\Phi_{0}$ is increased, coalescence is arrested at an earlier point. 

In addition to the benefit of increased stability, arrested coalescence
also constitutes a straightforward method for sculpting non-spherical
Pickering emulsion droplets. While these studies have successfully
shown how to control the resulting shape of the emulsion droplets,
little attention has been paid to the microstructure of the final
state, or the role of the arrest process itself. The purpose of this
paper is, therefore, to determine how the particle microstructure
is affected by the nonuniform curvature, and to disentangle the relative
influence of geometric influences and the kinetic process of relaxation
prior to arrest. 

Valuable insight into the role played by geometry comes from the rich
literature on packings on curved surfaces\citep{bowick2009two}, for
which spherical Pickering emulsion droplets---dubbed \emph{colloidosomes---}have
proven an ideal model system\citep{bausch2003grain}. Unlike 2D Euclidean
space where crystalline hexagonal packing is optimal, on curved surfaces
defects---particles with a contact number $c_{i}$ other than $6$---are
generally required to accommodate the curvature. Defining the defect
charge $q_{i}=c_{i}-6$, the total charge of the curved surface must
satisfy $\sum_{i}q_{i}=6\chi$, where $\chi$ is the Euler characteristic
of the surface.

Continuum elastic theory\citep{seung1988defects} can be used to analyze
the defect distribution on curved surfaces\citep{perez1997influence,bowick2000interacting,bowick2002crystalline,bowick2006crystalline,vitelli2006crystallography,giomi2007crystalline,giomi2008defective,irvine2010pleats}.
The free energy is dominated by the competition between the defect
core energy term, which favors fewer defects, and the compensating
elastic term that arranges the defect charge distribution to approximately
match the Gaussian curvature. For packings on spherical surfaces where
$\chi=2$, if the core energy is large, i.e. the number of particles
is small, there are only the twelve 5-fold disclinations required
by the Euler Theorem which form an icosahedron. Conversely, if the
core energy is small, and hence the number of particles is large,
excess dislocations appear with different signs to maintain the total
defect charge required by topology. The excess dislocations tend to
proliferate into grain boundaries, referred to as \emph{scars,} to
help screen the Gaussian curvature\citep{bowick2000interacting}.
Experiments have confirmed that the extra disclinations start to emerge
when the number of particles exceeds some limit\citep{bausch2003grain}.
Nonuniform curvature leads to localization of the defects\citep{irvine2010pleats,brojan2015wrinkling,jimenez2016curvature,burke2016jamming}. 

Insights from the packing literature are most applicable when the
particle structure is in quasistatic equilibrium with the host shape,
i.e. when relaxation of the droplets formed in arrested coalescence
occurs much more slowly than processes such diffusion that relax the
structure. We are, however, unaware of any prior attempt to understand
the role of kinetics in arrested coalescence. To that end, we simulate
as described in section \ref{sec:Model} the structures produced by
arrested coalescence using previous theoretical and experimental studies
of liquid drop coalescence\citep{studart2009arrested,pawar2011arrested,eggers1999coalescence,menchaca2001coalescence,wu2004scaling}.
We then separate the influence of geometry and kinetics on the microstructure
by comparing static packings produced with a fixed shape in section
\ref{sec:Geometry} and kinetic packings produced as the shape relaxes
towards the final spherical ground state in section \ref{sec:Kinetics}.
Properties of the surface that predict these effects are described
together with conclusions in section \ref{sec:Conclusion}. 

\section{Model\label{sec:Model}}

\begin{figure}
\begin{centering}
\includegraphics[width=1\columnwidth]{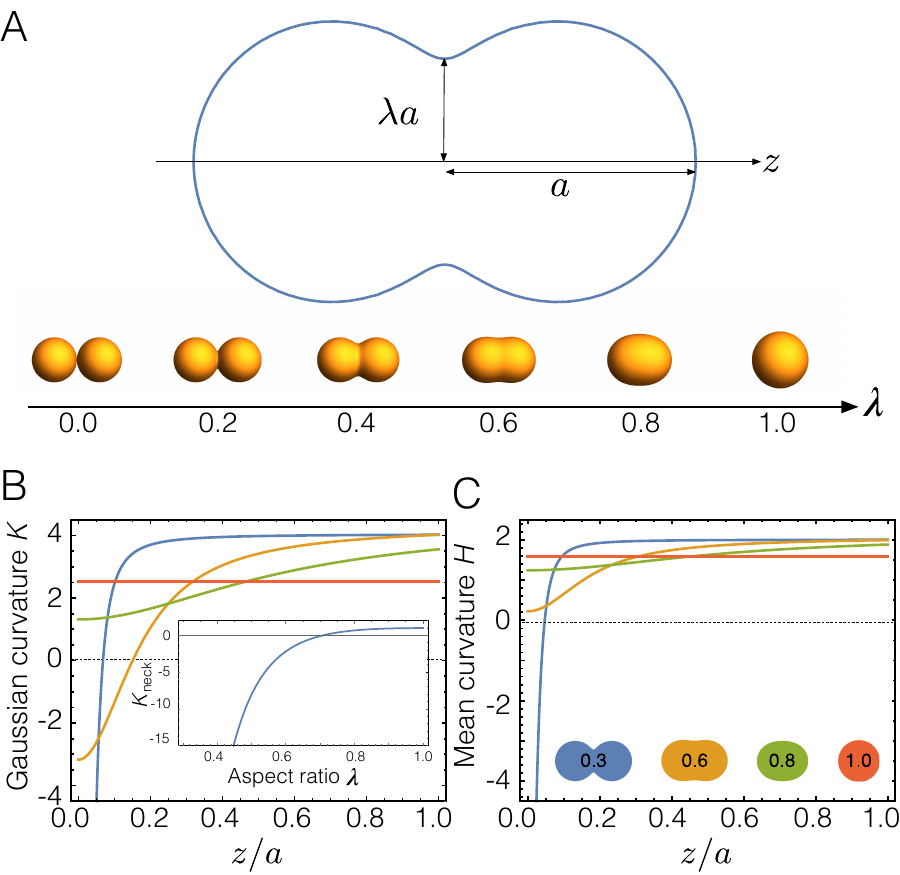}
\par\end{centering}
\centering{}\caption{\label{fig:droplet}\textbf{Ansatz fluid-fluid interface shape.} (A)
Simulation of different stages of coalescence parametrized by $\lambda$.
(B) The Gaussian curvature for several typical shapes of arrested
coalescence along the rotational symmetry axis with fixed volume.
Inset shows the curvature transition at neck $z=0$ for coalescence
evolution. (C) Corresponding plots of Mean curvature. }
\end{figure}

An analytical \emph{ansatz} to describe the shape evolution of a pair
of spheres as they relax following coalescence was proposed by Garabedian
\emph{et al.}\citep{garabedian2001model}. After initial contact,
the surface is described by the level set, 
\begin{equation}
\frac{a^{2}\lambda^{2}(x^{2}+y^{2})+a^{2}z^{2}}{(x^{2}+y^{2}+z^{2})^{2}}=1,
\end{equation}
where $a$ is the half length of long axis and aspect ratio $\lambda\in\left[0,1\right]$
controls the extent of coalescence as shown in Fig. \ref{fig:droplet}A.
The value $\lambda=0$ corresponds to the two droplets just touching
each other and $\lambda=1$ represents the final state as one spherical
droplet. The center is located at the origin and the $z$ axis is
the axis of rotational symmetry. The value $a$ is chosen for each
$\lambda$ to hold the total volume of the surface constant. 

The defect distribution in static packings is known to be controlled
by the distribution of Gaussian curvature, which acts like a nonuniform
background charge distribution in addition to the discrete defect
charges for the elastic term in free energy.\citep{giomi2008defective}.
We therefore display the Gaussian curvature along the axis of rotation
$z$ for several different values of $\lambda$ with fixed volume
as the sum of two spheres with diameter 1 in Fig. \ref{fig:droplet}B.
During the initial stages of relaxation, i.e. $\lambda\lesssim0.7$,
the neck of the doublet creates a region where the Gaussian curvature
$K$ is negative. We show in the inset of Fig. \ref{fig:droplet}B
the transition of curvature at the neck from negative to positive.
For $\lambda=0.3$, the curvature at the neck is large and negative
while at the ends of the droplet the curvature approaches a constant
value because here the surface is almost spherical. As $\lambda$
increases, the profile $K(z)$ becomes smoothed over time until it
approaches a uniform constant value for the spherical final state. 

Also shown in Fig. \ref{fig:droplet}C is the Mean curvature $H$,
which similarly exhibits a negative region at the neck for early stages
of the relaxation. This quantity is related to the capillary pressure
difference across the surface $\Delta p$ through the Young-Laplace
equation,
\begin{equation}
\Delta p=2\gamma H\label{eq:young}
\end{equation}
where $\gamma$ is the surface tension. As a measure, therefore, of
the local generalized force acting to evolve the shape towards the
equilibrium state, we shall show in section \ref{sec:Kinetics} that
the mean curvature plays a key role in the kinetically dominated regime. 

To simulate particles embedded on the surface we employ two different
algorithms fully described in the Methods section below. Static packings
are produced on shapes of fixed $\lambda$ using a Monte Carlo inflation
algorithm inspired by the Lubachevsky--Stillinger algorithm\citep{Lubachevsky1990};
we supplement this algorithm to ensure rigidity of the final packings.
A second algorithm, with particles of fixed radius $r$ that diffuse
with their centers of mass constrained to the evolving ansatz surface,
is used to resolve the influence of kinetics. While the particles
move by diffusion, the shape $\lambda(t)$ is slowly evolved at constant
volume such that the radius of the neck scales $\propto t^{1/2}$
with time. This form was proposed for the inertial regime where Reynolds
number is large by Eggers \emph{et al.}\citep{eggers1999coalescence}
and confirmed experimentally\citep{menchaca2001coalescence,wu2004scaling}
to hold in the early stages of coalescence. We use this power law
for the whole process for simplicity.

\section{Geometry\label{sec:Geometry}}

\begin{figure*}[th]
\begin{centering}
\includegraphics[width=7.2in]{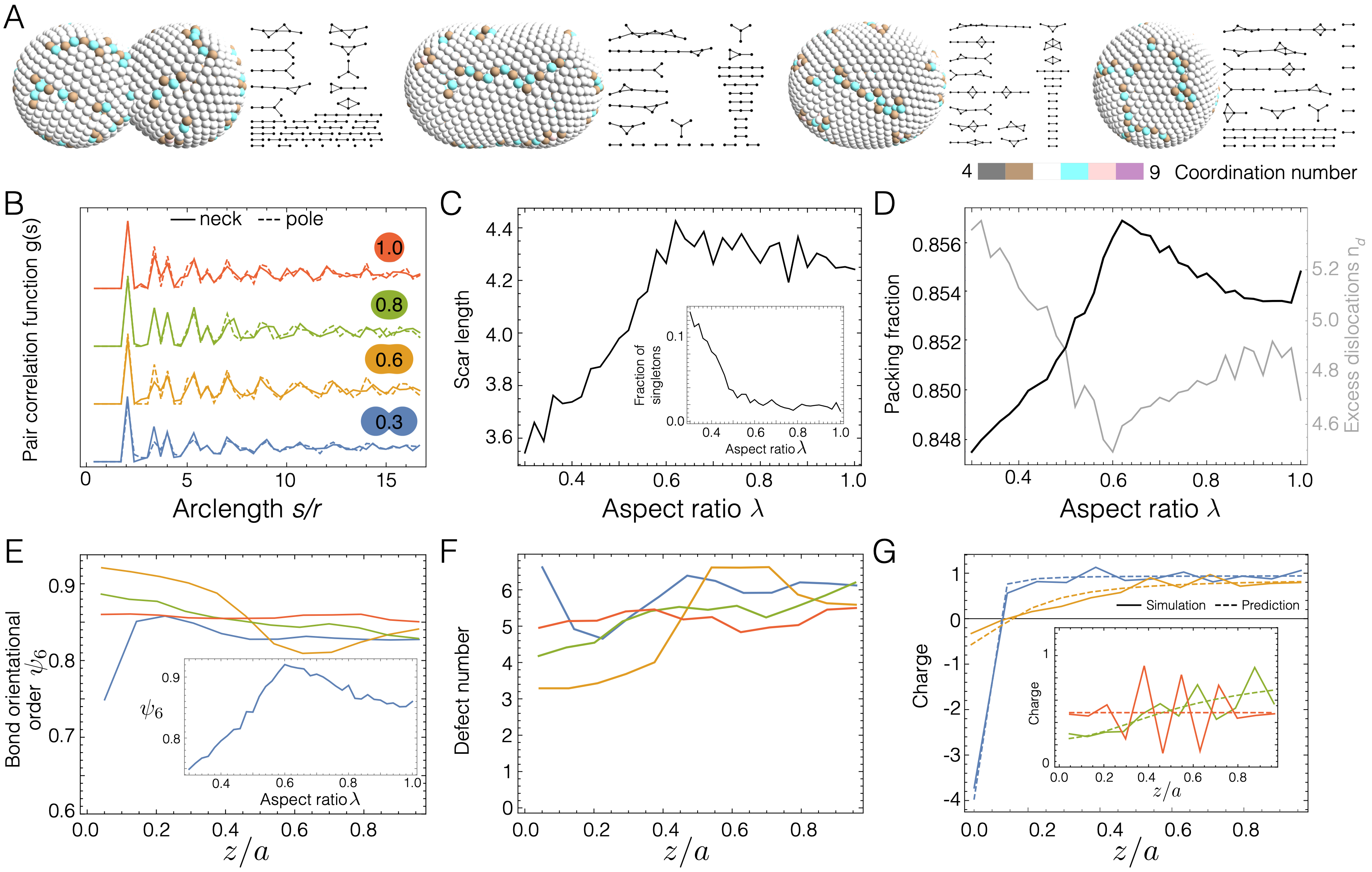}
\par\end{centering}
\caption{\label{fig:static}\textbf{Influence of geometry on the microstructure.}
(A) Representative packing configurations for $\lambda=0.3,0.6,0.8,1.0$
and their defect subgraphs. (B) Pair correlation function $g(s)$
as a function of arclength $s$ over particle radius $r$ for particles
around neck (solid lines) and pole (dashed lines). (C) Average scar
length as a function of $\lambda$ computed from defect subgraphs.
\emph{Inset: }Fraction of singletons as a function of $\lambda$.
(D) Packing fraction $\phi$ (black) and excess dislocations $n_{d}$
(grey) as a function of $\lambda$. (E) Bond orientational order parameter
$\psi_{6}$ distribution along the rotational symmetry axis of the
surface $z/a$. Inset shows the order parameter of the neck region
for different stages of coalescence. (F) Defect number distribution
along the rotational symmetry axis of the surface $z/a$. (G) Charge
distribution along the rotational symmetry axis of the surface $z/a$
from packing (solid lines) and integrated Gaussian curvature (dashed
lines) for $\lambda=0.3,0.6$ and (inset) $\lambda=0.8,1.0$.}
\end{figure*}

Using the protocol described above with fixed shape, we generated
100 rigid packings of 1000 particles for several values of the coalescence
parameter $\lambda\in[0.3,1]$. Representative results are shown in
Fig. \ref{fig:static}A, where the particles are colored by coordination
number computed from the Delaunay triangulation. Visually, the packings
appear largely crystalline with the expected chains of defects distributed
over the whole surface. Plots of the density-density pair correlation
function $g(s)$\citep{donev2005pair} as a function of arclength
$s$ in units of the particle radius $r$ for the particles around
the neck ($z=0$) and pole ($z=a$) are shown in Fig. \ref{fig:static}B
as dashed and solid lines respectively and appear similar. In agreement
with visual inspection, these plots show long range order and split
second peaks indicative of local crystalline order. 

To visualize the scar network, we compute the neighbor graph from
the Delaunay triangulation and delete all vertices with six-fold coordination\citep{Mascioli:2017ii}.
The disjoint components of this \emph{defect subgraph} correspond
to individual scars; typical scar networks are shown in Fig. \ref{fig:static}A
alongside the corresponding packing. The majority of the scars display
a linear morphology with some branching. No strong variation in scar
morphology is observed for different $\lambda$, however, the size
distribution of the scars does vary. In \ref{fig:static}A, the largest
scar gets longer as $\lambda\to1$. The scar length, here defined
as the mean number of vertices in the disjoint subgraphs, is plotted
in Fig. \ref{fig:static}C. For low $\lambda$ packings, shorter scars
are generated, and as shown in the inset of Fig. \ref{fig:static}C,
a higher fraction of them are isolated singleton defects. This is
because for low $\lambda$ the radius of the two spheres is smaller
and hence the effect of curvature is stronger. As the negative curvature
becomes less distinct, the scar length increases until $\lambda\thickapprox0.7$,
the point at which the region of negative Gaussian curvature disappears.
Beyond this value, the length of the scars doesn't increase further,
showing that positive Gaussian curvature generates similar scars. 

Consistent with prior work, we define $n_{d}$, the\emph{ }excess
dislocations beyond the twelve required by topology as\citep{burke2015role},
\begin{equation}
n_{d}=\frac{1}{2}\left(\frac{\sum_{i}|q_{i}|}{12}-1\right),
\end{equation}
where the sum is over dislocations and $q_{i}$ represents the charge
of the $i$th dislocation, and display $n_{d}$ as a function of $\lambda$
in Fig. \ref{fig:static}D. For low $\lambda$, corresponding to early
arrest, the large amount of negative curvature induces more dislocations.
As the neck region becomes flatter, the defect number decreases and
reaches a minimum at $\lambda=0.6$. For $\lambda>0.6$, more dislocations
emerge to accommodate the positive curvature until the final spherical
stage. The packing fraction $\phi$, defined as the fraction of the
surface covered by particles, is also shown in Fig. \ref{fig:static}D
and follows an inverted trend because the emergence of dislocations
reduces the packing fraction. Since the shape away from the neck is
spherical, this variation in the $n_{d}$ appears mainly due to the
curvature distribution around the neck. 

To test this, spatially resolved plots of the bond orientational order
parameter $\psi_{6}=\left\langle \exp(i6\theta)\right\rangle $\citep{nelson1979dislocation}
for several $\lambda$ are displayed in Fig. \ref{fig:static}E. In
these plots, the surface is divided into 24 equal-area axially symmetric
regions and the symmetry of the shape is exploited by collapsing corresponding
regions for positive and negative $z$. The orientational order is
generally quite high, greater than $0.8$ everywhere except close
to the neck for $\lambda=0.3$, and as expected is spatially uniform
for the spherical case $\lambda=1$. The distribution of $\psi_{6}$
undergoes a transition around the neck area: for $\lambda=0.3$, the
region close to the neck has low $\psi_{6}$, with a local maximum
at around $z/a=0.2$. In contrast, for $\lambda=0.6$ the neck region
has enhanced $\psi_{6}$, and a less ordered region away from the
neck. As $\lambda$ increases further, the amplitude of this variation
is reduced, with lower order around the neck.

In the inset of Fig. \ref{fig:static}E, the mean value of $\psi_{6}$
at the innermost region is shown as a function of $\lambda$. As $\lambda$
increases, the neck becomes less pronounced, and the order increases;
after $\lambda\sim0.6$, when the neck becomes flat, the appearance
of the positive curvature reduces the order again. 

Consistent with this, the number density of the defects, shown in
Fig. \ref{fig:static}F, closely resembles the inverted form of $\psi_{6}$
since defects reduce the bond orientational order dramatically. More
defects are induced by the large negative curvature around the neck
while the flat region has fewer defects and the number of defects
increases again as the neck gains positive curvature. As $\lambda\to1.0$,
a uniform distribution of defects occurs due to the uniform curvature. 

Finally, in Fig. \ref{fig:static}G and its inset, the defect charge
density is shown in solid lines, together with a predicted distribution
from the integrated Gaussian curvature in dashed lines. The defect
charge distribution is well explained by the curvature distribution
of the surface and also the constraint that the total charge must
be $12$ from the Euler characteristic of the surface. For $\lambda=0.3$,
strongly negative Gaussian curvature at the neck as apparent in Fig.
\ref{fig:droplet}B tends to induce negative defects. Hence, additional
positive charges must be generated at the ends to satisfy the topological
constraint. Conversely, for $\lambda=0.6$, fewer negative defects
are induced and the neck region has almost zero mean curvature leading
to an overall enhancement of the order. Defects are needed outside
the neck area to satisfy the charge constraint and also to match the
positive curvature. 

This section has elucidated the role played by the nonuniform geometry
on the final states observed in arrested coalescence, notably the
key role that the neck plays in determining the distribution of defects.
For shapes corresponding to early arrest, when the neck possesses
strong negative curvature, additional defects are induced, while at
shapes corresponding to intermediate arrest, the flat neck region
promotes locally enhanced order. Scars at early arrest are shorter,
because the radius of curvature at the caps is smaller; the scar length
grows until the region of negative curvature vanishes. Consistent
with previous work, the defect charge distribution is well predicted
by the distribution of Gaussian curvature and the overall charge constraint
imposed by topology. 

\section{Kinetics\label{sec:Kinetics}}

\begin{figure}
\begin{centering}
\includegraphics[width=3.5in]{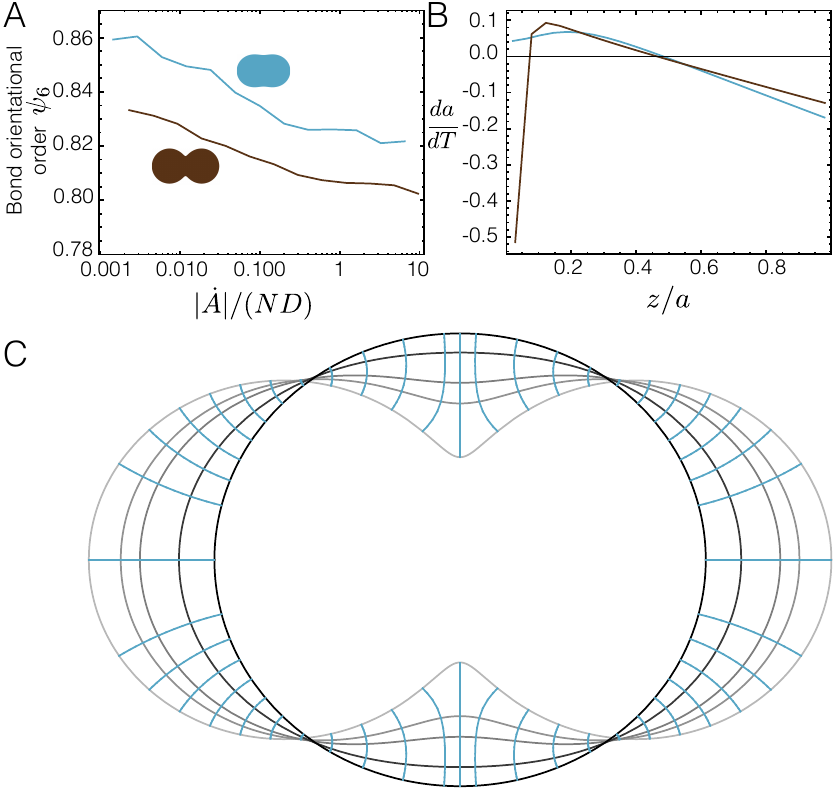}
\par\end{centering}
\caption{\label{fig:dynamicTotal}\textbf{Evolution of particles on the evolving
surface. }(A) Bond orientational order parameter $\psi_{6}$ of all
particles as a function of $\frac{|\dot{A}|}{ND}$. Brown is for system
arrested at $\lambda\sim0.3$ and blue at $\lambda\sim0.6$. (B) Corresponding
rate of area change as a function of position. (C) Shape evolution
of the surface (gray lines) with geodesics followed by fixed particles
(blue). }
\end{figure}

We now examine the role played by kinetics on the arrested state.
A natural measure of the importance of kinetics is the rate of total
area change $\dot{A}$ at the arrest point, which may be compared
to the diffusion coefficient $D$ because they have the same dimensions.
We therefore construct a new dimensionless parameter,
\begin{equation}
\frac{\dot{A}}{D}=\frac{dA}{dt}\frac{\tau_{d}}{2r^{2}}=\frac{1}{2r^{2}}\frac{\tau_{d}}{\tau_{r}}\frac{dA}{dT}\label{eq:dynamic parameter}
\end{equation}
and nondimensionalize time using $\tau_{d}$ the corresponding diffusion
timescale defined through $2r=\sqrt{2D\tau_{d}}$, $T$ is the dimensionless
time $t/\tau_{r}$ and $\tau_{r}$ the time of full relaxation. The
ratio 
\begin{equation}
\theta=\tau_{d}/\tau_{r}\label{eq:theta}
\end{equation}
therefore also emerges as a measure of the relative influence of diffusion
and relaxation and is in practice the independent variable that is
varied to construct our ensemble of simulations. We shall also make
use of the quantity $\dot{A}/(ND)$ where $N$ is the number of particles
to quantify the average rate of area change per particle, together
with a local version $\frac{\dot{a}}{nD}$, where $a$ is the area
of some region of interest on the surface and $n$ is the number of
particles in that region. The quantity $dA/dT$ is almost linearly
related to $\lambda$ for all but small values of $\lambda<0.1$ where
the \textit{ansatz} is a poor approximation to the experiment\citep{pawar2011arrested},
and so we do not investigate arrest in this regime.

Using the protocol described in section \ref{sec:Model}, a set of
arrested states was generated with $N=800$ particles and suitable
particle radius to promote a point of arrest $\lambda_{a}$ for two
different scenarios: an early arrest case with $\lambda_{a}\approx0.3$,
where the curvature of the neck is extremely negative and late arrest
at $\lambda_{a}\approx0.6$, where the neck is flat. The ratio of
diffusion time scale to the total relaxation time $\theta=\tau_{d}/\tau_{r}$
is varied from $2^{-8}$ to $2^{4}$ with 50 samples for each value. 

In Fig. \ref{fig:dynamicTotal}A, the bond orientational order parameter
$\psi_{6}$ is shown as a function of $|\dot{A}|/ND$, which is, as
was shown in Eq. \ref{eq:dynamic parameter}, linearly related to
$\theta$. This choice of parameter accounts for the fact that the
rate of area change is naturally larger for early arrest. Faster relaxation
leads, as might be expected, to less ordered structures for both scenarios,
and overall the order is higher for late arrest. This latter is consistent
with the results of Fig. \ref{fig:static}E that show more ordered
packings for the $\lambda=0.6$ surface and is therefore geometric
in origin. 

Before examining microstructural differences between the two scenarios,
we pause to discuss the effect of the shape evolution on the particle
motion. A single free particle at rest on the initial surface moves
only subject to constraint forces which lie along the direction locally
normal to the surface. We neglect the inertia of the particles which
is assumed to be damped by the surrounding fluid. Representative trajectories
are visualized in Fig. \ref{fig:dynamicTotal}C for several starting
positions around the doublet. In some locations, for example the caps,
the trajectories are convergent indicating that particles here tend
to be compressed as the doublet relaxes; elsewhere the trajectories
are weakly divergent indicating that here particles would be pushed
apart. 

We can use a similar approach to define a parametrization-independent
local rate of change in area $\dot{a}$ by dividing the initial surface
into annuli of equal area and evolving the boundaries along the normal
direction. The instantaneous $\frac{da}{dT}$ as a function of position
is plotted in Fig. \ref{fig:dynamicTotal}B for the early and late
arrest scenarios, where $T$ is the dimensionless time $t/\tau_{r}$.
An important difference is that for early arrest $\lambda_{a}\approx0.3$,
a very strong region of compression exists close to the neck; this
is missing for $\lambda_{a}\approx0.6$ where particles at the neck
are being pushed apart by the constraint forces. Compressive regions
exist in both scenarios around the caps. 

Identical particles initially uniformly distributed on the doublet
therefore evolve differently depending on their position. Those near
the caps are pushed together and should crowd first, suggesting that
the crystalline region should develop first at the caps and grow,
reminiscent of the growth fronts observed in \citep{Waitukaitis:2012dv,Han:2016bj}.
If the arrest is early, crowding may also occur around the neck. Particles
in the intermediate region between the neck and the cap have more
opportunity to relax than particles in the compressive regions. 

\begin{figure}
\begin{centering}
\includegraphics[width=1\columnwidth]{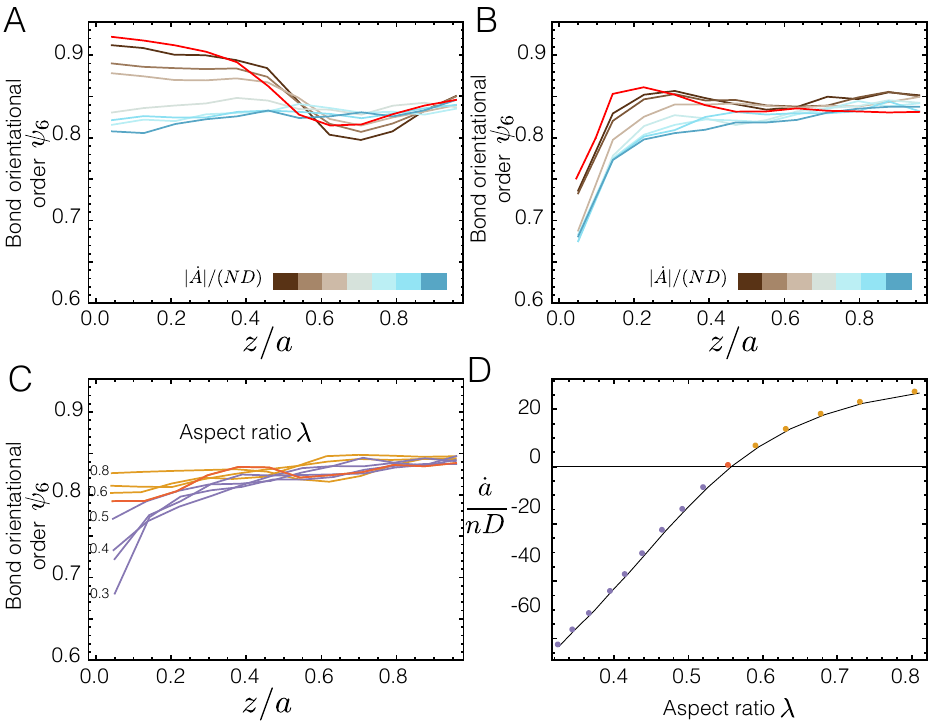}
\par\end{centering}
\caption{\label{fig:dynamicLocal}\textbf{Influence of dynamics on the microstructure.
}(A) Bond orientational order parameter $\psi_{6}$ distribution along
the surface for late arrest $\lambda_{a}\approx0.6$ at different
values of the dynamical parameter $|\dot{A}|/ND=(0.0028,0.012,0.048,0.20,0.39,1.58,6.36)$.
The red line is the distribution for static packing. (B) Corresponding
plot for $\lambda_{a}\approx0.3$ with $|\dot{A}|/ND=(0.0023,0.0091,0.037,0.15,0.59,2.38,9.5)$.
(C) $\psi_{6}$ distributions as a function of the aspect ratio of
the arrest point for fast relaxation $|\dot{A}|/ND\sim10$. (D) Local
rate of area change per particle at the neck as a function of $\lambda$
for fast relaxation.}
\end{figure}

The spatially resolved bond orientational order parameter is shown
for both scenarios in Fig. \ref{fig:dynamicLocal}A and B for $\lambda\approx0.6$
and $\lambda\approx0.3$ respectively. Different traces correspond
to different values of $|\dot{A}|/ND$ and, as before, the left and
right portions of the doublet are combined into one plot. In both
cases, the static order parameter distribution, shown by the red line,
is recovered for sufficiently slow relaxation. For late arrest, this
corresponds to an enhancement of $\psi_{6}$ in the flatter central
area and reduced order closer to the cap. As $|\dot{A}|/ND$ increases,
these features are softened and as $|\dot{A}|/ND\gtrsim1$, the orientational
order converges on a uniform constant value of $\sim0.83$\lyxadded{Timothy Atherton}{Mon Feb 18 16:11:52 2019}{
}which is the minimum order achieved by a static packing with similar
shape.

In contrast, for early arrest, the distribution of $\psi_{6}$ remains
similar as a function of $|\dot{A}|/ND$: there is a strong reduction
in the order parameter around the neck and a uniform distribution
near the cap. Increasing $|\dot{A}|/ND$ reduces the order globally,
shifting the curves down by as much as $0.1$ but does not change
their overall form. 

To understand the role of the arrest point more carefully, an ensemble
of simulations with different arrest points $\lambda$ was run in
the kinetically dominated regime $|\dot{A}|/ND\approx10$ and the
resulting distributions of $\psi_{6}$ are shown in Fig. \ref{fig:dynamicLocal}C.
Comparing these with equivalent plots for static packings, Fig. \ref{fig:static}E,
we see that for arrest earlier than $\lambda\sim0.5$, the static
and dynamic order distributions appear similar, although the overall
values of $\psi_{6}$ are attenuated. For arrest after $\lambda\sim0.5$,
the distributions no longer resemble the static distributions, instead
having approximately uniform $\psi_{6}\approx0.83$. Rapid relaxation
therefore appears more readily able to wash out variations caused
by the nonuniform curvature for late arrest. 

The transition in behavior appears to coincide with the point at which
the compressive region disappears from the neck, as can be seen in
Fig. \ref{fig:dynamicLocal}D which displays the instantaneous local
dynamical parameter $\dot{a}/nD$ at the neck at the point of arrest.
For early arrest, therefore, the signature of Gaussian curvature remains
imprinted on the microstructure regardless of the overall rate of
relaxation. 

\begin{figure}
\begin{centering}
\includegraphics[width=1\columnwidth]{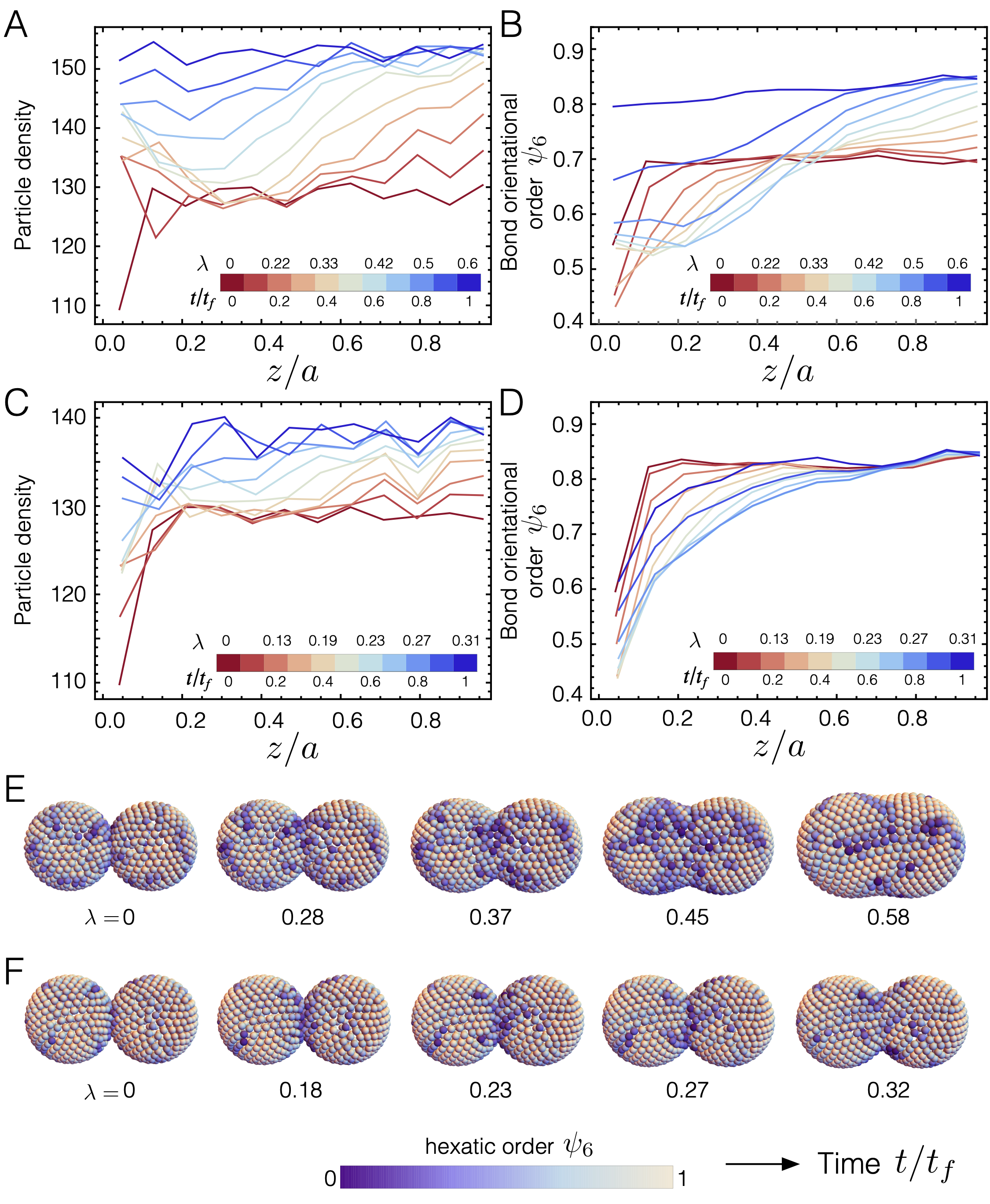}
\par\end{centering}
\caption{\label{fig:dynamictimeresolved}\textbf{Time resolved evolution of
the microstructure for the kinetically dominated regime.} Distributions
of (A) particle number density and (B) orientational order parameter
$\psi_{6}$ as a function of time for late arrest final aspect ratio
$\lambda_{a}\approx0.6$ and fast relaxation $|\dot{A}|/ND=9.5$.
(C) and (D) Corresponding plots for early arrest $\lambda_{a}\approx0.3$.
(E) and (F) Representative visualizations for $\lambda_{a}\approx0.6$
and $\lambda_{a}\approx0.3$ respectively; particles are colored by
$\psi_{6}$.}
\end{figure}

Time resolved analysis of the orientational order parameter allows
us to understand the transition further. The particle number density
is plotted at several time points $t/t_{f}$ in Fig. \ref{fig:dynamictimeresolved}A
for the late arrest scenario at $\lambda_{a}\approx0.6$, showing
growth of a dense region from the cap and neck as predicted from the
geometric argument above. Corresponding plots of the the $\psi_{6}$
distribution are displayed in Fig. \ref{fig:dynamictimeresolved}B,
and some simulation snapshots are visualized in Fig. \ref{fig:dynamictimeresolved}E.
Here $t_{f}$ is defined to be the time at which arrest occurs. Corresponding
plots and visualizations for early arrest $\lambda_{a}\approx0.3$
are displayed in Fig. \ref{fig:dynamictimeresolved}C, D and F respectively.
Neighbors of particles are not determined from Delaunay triangulation
since the particles are generally not densely packed during the coalescence.
Instead, we define neighbors as particles within the center-to-center
distance of $1.5$ times diameter. In practice, this definition changes
the numerical values of $\psi_{6}$ very little and doesn't affect
the trend of evolution. 

For late arrest particles are initially uniformly ordered, except
close to the neck. As relaxation proceeds, compression at the cap
produces a denser region that converges to about $\psi_{6}\sim0.83$.
This value is similar to the ordering observed in the static packing
at the cap. At the neck, the particles are initially more disordered
because of the strong concentration of negative Gaussian curvature,
and the initial compressive region that exists here begins to widen
the disordered region. At around $\lambda\sim0.4$, however, the widening
stops, and as $\lambda$ increases further the ordered region growing
in from the caps completely overcomes the disordered neck region. 

For early arrest, the ordering at the cap changes very little through
the course of the simulation, remaining similar to the static value.
The initially disordered region at the neck begins to widen, as it
did for the late arrest scenario, and reaches a maximum width at around
$\lambda=0.25$. After this, the neck region begins to uniformize
but does not completely do so, freezing in the disordered region that
is also seen in the static case. 

The influence of kinetics on the microstructure therefore depends
critically on the arrest point: the history of compression and expansion
of the surface may freeze in disordered regions caused by the underlying
curvature, even where relaxation might be expected to wash out the
effect of geometry. For early arrest, the strong initial compression
leads to a disordered region that is frozen in at all relaxation speeds
tested, while for late arrest fast relaxation fully blurs out the
nonuniform distribution of order promoted by geometry. 

\section{Discussion}

The static packings produced in section \ref{sec:Geometry} reconfirmed
the well-known role of the Gaussian curvature in determining the distribution
of order. The role of kinetics was shown in section \ref{sec:Kinetics}
to be surprisingly complicated, with a disordered region present in
the neck for early arrest, while for late arrest kinetics is able
to fully wash out variations in the distribution of order caused by
geometry. The final microstructure therefore depends critically upon
the history of compressive and and expansive regions as the relaxation
proceeds. These regions are identified by examining local contributions
to the the rate of change of the area $\dot{A}$. 

We can compute $\dot{A}$ from more fundamental quantities as follows.
Suppose we have a one-parameter family of surfaces $\mathbf{X}(\lambda)$
that describes the shape evolution of the surface. The rate of change
of area can then be written, 
\begin{equation}
\dot{A}=\int\nabla\cdot\mathbf{N}\left(\frac{d\mathbf{X}}{d\lambda}\cdot\mathbf{N}\right)\frac{d\lambda}{dt}dA,\label{eq:kinetic}
\end{equation}
where $\mathbf{N}$ is the local outward surface normal and the integral
is over the surface. The first factor is the divergence of the normal
and can be rewritten as the mean curvature, 

\begin{equation}
\dot{A}=\int2H\left(\frac{d\mathbf{X}}{d\lambda}\cdot\mathbf{N}\right)\frac{d\lambda}{dt}dA,\label{eq:kinetic2}
\end{equation}
while the second factor in (\ref{eq:kinetic}) and (\ref{eq:kinetic2})
is the normal component of the velocity as the surface evolves according
to $\lambda(t)$. We identify the integrand of (\ref{eq:kinetic2})
as $\dot{a}$. We note that (\ref{eq:kinetic2}) reveals an elegant
distinction: the microstructure of the static packings is determined
by the Gaussian curvature, an \emph{intrinsic} quantity, while the
role of kinetics is determined by \emph{extrinsic} quantities---those
that depend on how the surface is embedded---such as the mean curvature. 

The discriminant $\text{sign}(\dot{a}/D)$ that identifies regions
of compressed and expansion furthermore includes both properties of
the surface at arrest and quantities that depend on its evolution.
Importantly, the discriminant arises \emph{both} from the sign of
the mean curvature \emph{and} whether the motion is locally inward
or outward. At the caps, $H$ is positive but the motion is inward;
conversely at the neck $H$ is negative, at least for low values of
$\lambda$, but the motion is outward. One can therefore imagine manipulating,
through a judicious choice of interface, both of these quantities
to exploit kinetics as a means to control the microstructure. We note
that the evidence of the previous section suggests that, at least
crudely, it is the value of these quantities immediately prior to
arrest that are most important for determining the microstructure:
for the late arrest scenario $\lambda_{a}\sim0.6$, the compressive
regions that exist early in the relaxation appear to have little influence
on the final state. 

\section{Conclusion\label{sec:Conclusion}}

This work aimed to distinguish the relative contribution of geometry
and kinetics in determining the microstructure of droplets produced
by arrested coalescence. Relaxation of the shape drives the transition
to a final rigid structure, while motion of the particles, here modelled
as diffusive, facilitates reordering of the structure to achieve higher
packing fraction. The importance of each processes was shown to be
quantified by a single parameter $\dot{A}/ND$ which compares the
rate of change of area experienced by each particle to its diffusion
constant. As $\dot{A}/ND\to0$, the microstructure closely resembles
the static packing scenario where the system distributes defects to
match the Gaussian curvature. At finite $\dot{A}/ND$, kinetics tends
to reduce the overall order of the arrested structure, and tends to
blur out modulation of the order as the structure can no longer be
fully relaxed. Disordered regions induced by the Gaussian curvature
can, however, remain even in the kinetically dominated regime because,
as the surface relaxes, regions of compression and expansion serve
to trap them. The kinetic influence therefore, perhaps surprisingly,
depends on geometric properties of the hypersurface including mean
curvature but these are extrinsic in origin, i.e. depending on the
embedding. 

The remarkably rich influence of kinetics suggests the possibility
of exploiting it as a means to control the microstructure of colloidosomes.
One can speculatively envisage designing a shape using the formula
(\ref{eq:kinetic2}) that incorporates compressive regions prior to
arrest and could selectively lock in disordered regions; these might
become targets for further coalescnce events in multistage assembly.
The extent to which one is free to choose the mean curvature and the
local velocity of the surface remains an open problem, however. The
Young-Laplace equation Eq. (\ref{eq:young}) implies that locally
these are not independent quantities. That they are different in the
present work shows that constraints such as the overall volume play
an essential role, and the consequent restrictions on the design space
remain to be illuminated. 

A second direction that should be pursued is the connection of these
arrested shapes to \emph{jamming}, a transition to rigidity as a function
of density that occurs in particulate media\citep{torquato2010jammed,liu2010jamming}.
While the kinetically arrested structures observed here are \emph{not}
jammed, in that they may possess unconstrained collective motions
of particle, the static packings we use as a comparison \emph{are}
because this is explicitly enforced. Two of the present authors recently
proposed that rigid structures formed as a result of shape evolution
form a new class called\emph{ ``metric jamming''} \citep{burke2016jamming}
where the final state is rigid both with respect to perturbations
of the particles and the manifold on which they are embedded. Analysis
of the arrested coalescence problem along these lines may help determine
the longevity of the undoubtedly metastable arrested structures, as
well as provide tools to determine the mechanical properties of these
structures. 

\section{Methods\label{sec:Methods}}

\textbf{Static packings---}Particles are initially dispersed with
their center of mass on the surface at zero radius, diffused by Brownian
motion according to the Langevin equation,

\begin{equation}
\mathbf{x}_{i}^{'}(t+\varDelta t_{p})=\mathbf{x}_{i}(t)+\mathbf{\eta}_{i}\sqrt{2D\Delta t_{p}},\label{eq:langevin}
\end{equation}
where $\mathbf{\eta}_{i}$ is a random step drawn from Gaussian distribution
along the tangent plane, and $D$ is the diffusion constant such that
the variance of stepsize for Brownian motion in time $t$ is $2Dt$.
We may therefore define a characteristic diffusion time scale $\tau_{d}$
that gives a standard deviation of stepsize equal to the particle
diameter, 
\begin{equation}
2r=\sqrt{2D\tau_{d}}.\label{eq:diffusionTimeScale}
\end{equation}
As the particles diffuse, their radii $r$ are increased (inflation
moves) very slowly, with $\frac{\delta r}{\sqrt{2Dt}}\sim10^{-4}$
in unit time. Collective motions that undo overlap are found at each
stage by gradient descent on an artificial potential, 

\begin{equation}
V_{overlap}=\begin{cases}
\begin{array}{cc}
r^{2}-rx & x<r\\
0 & x\geq r
\end{array},\end{cases}\label{eq:Voverlap}
\end{equation}
that penalizes overlap. The simulation is halted when no further move
is possible without inducing overlaps. 

Generically, packings produced by this algorithm need not be rigid,
i.e. there may exist collective motions of particles that can unjam
the system and allow further relaxation of the surface. We therefore
adapt\citep{burke2016jamming} a linear programming approach \citep{donev2004linear}
to identify these collective motions, execute them, and restart the
packing simulation. Before applying the linear program, the configuration
is conditioned by minimizing an artificial soft repulsive potential
imposed between all pairs of particles; this tends to push the particles
away from one another. This process is repeated until a rigid final
state is achieved. 

\textbf{Dynamic simulations---}A second algorithm was used to understand
how the relaxation process affects the final structure. For these
simulations, particles are initially dispersed by random sequential
deposition with a fixed particle radius $r$. During the simulation,
diffusion moves are made as before in Eq. (\ref{eq:langevin}). During
relaxation moves, the particles are constrained to the surface with
overlaps prevented to first order using Lagrange multipliers\citep{vesely2013pendulums}.
After each relaxation step, Eq. (\ref{eq:Voverlap}) is minimized
to remove overlaps. If not all overlaps could be undone, the timestep
is reduced. The algorithm halts when the timestep of relaxation is
smaller than a threshold $\delta t$. 
\begin{acknowledgments}
This material is based upon work supported by the National Science
Foundation under Grant No. DMR-1654283.
\end{acknowledgments}

\providecommand*{\mcitethebibliography}{\thebibliography}
\csname @ifundefined\endcsname{endmcitethebibliography}
{\let\endmcitethebibliography\endthebibliography}{}

\end{document}